\documentclass{article}
\usepackage{amsmath,amsthm,amssymb}

\begin{document}
\title{The Eight Vertex Model.New results.
 \footnote{Talk with title ``An Elliptic Current Operator for the Eight Vertex  Model''
presented at the Solvay workshop BETHE ANSATZ: 75 YEARS LATER, October 19-21, 2006,
 Brussels,Belgium}}
\author{Klaus Fabricius\footnote{email: Fabricius@theorie.physik.uni-wuppertal.de},\\
        Fachbereich C, Bergische Universit\"at Wuppertal,
        D-42097 Wuppertal\\
       Barry M. McCoy\footnote{email: mccoy@insti.physics.sunysb.edu} \\
Institute for Theoretical Physics, State University of New York,\\
Stony Brook,  NY 11794-3840}
\maketitle
\begin{abstract}
\noindent
Whereas the tools to determine the eigenvalues of the eight-vertex
transfer matrix $T$ are well known there has been until recently
incomplete knowledge about the eigenvectors of $T$. We describe the 
construction of eigenvectors of $T$ corresponding to degenerate eigenvalues
and discuss the related hidden elliptic symmetry.
\end{abstract}
\section{Introduction}
\label{intro}
\noindent
The eight vertex model was introduced into physics by Baxter
\cite{bax72} as the
natural generalization of the six vertex model which satisfies the
star triangle equations and because the six vertex model is solved by
the same ansatz for eigenvectors which was introduced by 
Bethe \cite{bethe} in his
famous 1931 computation of the eigenvectors and eigenvalues of the
isotropic Heisenberg chain  it is a common practice of many authors to
present the computation of the eigenvalues and eigenvectors of the
eight vertex model as a generalization of the methods of Bethe.
Indeed it is for this reason that the 8 vertex model is the topic of
many talks at this conference honoring the 75 anniversary of Bethe's
original paper.

Nevertheless there are several profound distinctions between the 6 and 8
vertex model which make the methods used to study the properties 
of the 8 vertex model very different from  the methods used in the
Bethe's ansatz solution of the 6 vertex model as presented, for
example, in the classic 1967 papers by Lieb \cite{Lieb} 
and by Sutherland, Yang and Yang \cite{syy}  
and the corresponding 1966 papers of Yang and  Yang \cite{yy}
for the anisotropic Heisenberg model. Perhaps the most visible of
these differences in solution is that while the Bethe ansatz for the 
eigenvectors of the 6 vertex model is valid for all values of the
crossing parameter $\eta$ (and the corresponding anisotropy $\Delta$
of the anisotropic Heisenberg chain) a very important role for the
solutions of the 8 vertex model given in \cite{bax72} is played 
by what is called the ``root of unity'' condition
\begin{equation}
2L\eta=2m_1K+im_2K'
\label{root1}
\end{equation}
where $2K$ and $2K'$ are the real and imaginary periods of the
quasiperiodic functions which characterize the Boltzmann weights of
the 8 vertex model. In particular the condition (\ref{root1}) is necessary 
for the original  solution of the
eigenvalue problem of the 8 vertex model transfer matrix of 
\cite{bax72} to hold and for  the solution for eigenvectors given in
\cite{bax731}--\cite{bax733} to hold it is necessary that 
the related root of unity condition
\begin{equation}
L\eta=2m_1K+im_2K'
\label{root2}
\end{equation}
be imposed. 
A study of the eigenvalue problem for generic values of
$\eta$ is given in \cite{bax731} and \cite{baxb}  
however none of these
classic studies either for roots of unity or for generic values
revealed all the properties of the eigenvalue spectrum.
In particular it was found in \cite{Newdev} that the 
condition (\ref{root1}) is  not sufficient and 
for the method of \cite{bax72} to hold we must require in
(\ref{root1}) with $m_2=0$ that
for a system with $N$ lattice sites in a chain that  
the case with $L$ odd and $m_1$ even must be excluded for $N>L-1$ if $N$ is
even and for all $N$ if $N$ is odd.

The eigenvectors 
at generic values of $\eta$ are briefly mentioned 
in \cite{TakFadd},\cite{FelderVarch}, and \cite{bax2002} 
and in the recent paper of Bazhanov and
Mangazaeev \cite{BazMang} presented at this conference. 

It is the purpose of this paper to present recent progress in the
study of the eigenvalues and eigenvectors of the 8 vertex model
at the roots of unity (\ref{root1}) with $m_2=0$ and $L$ is  even or 
both   $L$ and $m_1$ are odd for even values of $N$.
We found in \cite{Newdev} that
one has to distinguish two types of eigenstates of $T$:\\
1. singlet states for the nondegenerate eigenvalues.\\
2. states appearing in multiplets for degenerate eigenvalues.\\
The singlet states are eigenstates of the spin reflection operator.
 This should be compared with the limiting case of the 
six vertex model where only a
few states with $S^z=0$ are singlets. Furthermore in the Bethe 
ansatz treatment of \cite{bethe}-\cite{yy} 
the eigenvectors  of the Hamiltonian are by definition eigenvectors of 
the operator $S^z$ and thus for  $S^z\neq 0$ none of the
Bethe states can be an eigenstate of the spin reflection operator.

Little attention has usually been paid to the second case
despite the fact that depending on the size $N$ and the 
value of the crossing parameter
$\eta$  the overwhelming majority of states is degenerate.
As an illustration we mention that for $N=12$ and $\eta=K/2$ there are
128 non degenerate states in the eigenspace of dimension 4096.
This same phenomenon of degeneracies at roots of unity 
which are larger than the symmetry
required under $S^z\rightarrow -S^z$ occurs in the six-vertex model
where it is explained by the existence of an $sl_2$ loop algebra
\cite{dfm}. The explanation of the
 multiplets as highest weight representations of this algebra
based on the algebraic Bethe ansatz of \cite{TakFadd}
was first given in \cite{Odyssey} and a rigorous proof in terms of
representation theory is given in \cite{deg4}.
A treatment  in terms of the coordinate Bethe ansatz is 
given in \cite{bax2002}.

The more complicated problem of obtaining all degenerate eigenstates 
of $T$ in the algebraic 
Bethe-Ansatz  of the eight-vertex model has been solved recently 
in \cite{elliptic}.
A parallel investigation for the description of the eight-vertex model 
given in \cite{FelderVarch}
is done in \cite{deg2,deg3}.

\section{Degenerate eigenvectors of $T$ in the eight-vertex model.}
\noindent
Our goal is to find a general procedure allowing the construction of
all degenerate eigenstates of the transfer matrix $T$.
The powerful and elegant algebraic Bethe-Ansatz \cite{TakFadd} is 
the ideal tool 
to develop the analytic framework in a transparent manner. Furthermore it
will allow the interpretation of the result in the context of 
symmetry algebras.\\
We shall first describe the important results obtained in
\cite{TakFadd}. For the six vertex model
the authors find that eigenstates of $T$ are given by a 
product of $B$-operators acting
on a reference state. In the six-vertex model this works for all irrational 
crossing parameters.
For rational multiples of $\pi$ not all eigenstates
are obtained but only those which are highest weight states of loop algebra
multiplets. To construct the remaining elements of these multiplets in addition
to $B$-operators the more complicated creation operators of strings are needed.
These creation operators of strings were introduced in \cite{Odyssey}.
Besides being creation operators of strings they have an important 
algebraic meaning: They are
current operators of the loop algebra symmetry of the six-vertex model.
The description of the loop-algebra symmetry of the six-vertex 
model is given in \cite{dfm}.\\
In the eight-vertex model the situation is more complicated. 
First the well developed
results \cite{bax733,TakFadd} are valid only at roots of unity. 
In \cite{TakFadd} eigenvectors of $T$ are given by finite
sums of products of $B$-operators acting on generalized reference states.
Like in the six-vertex model one gets by this method 
all singlet states. But unlike  the  case of the
six-vertex model the
$B$-operators and the reference states of the eight vertex model 
depend on additional 
free parameters $s,t$ and by varying these parameters
one gets a subspace of dimension larger than 
one for each degenerate eigenspace.
However there still remain a large number of states states in the multiplets 
which cannot be constructed
in this way. For the construction of these missing eigenstates  one
needs the additional 
string creation operators which were found in \cite{elliptic}.
There are also missing eigenvectors in the solution of the 
eight-vertex model by
Felder and Varchenko \cite{FelderVarch}. They have been studied 
by Deguchi in \cite{deg2,deg3}.

\section{The dimension of degenerate subspaces in the eight-vertex model.}
\noindent
The transfer matrix $T(v)$ has the property that it satisfies
the famous $TQ$ equation derived in \cite{bax72} 
\begin{equation}
 T(v)Q(v)=[\rho h(v-\eta)]^NQ(v+2\eta) +[\rho h(v+\eta)]^NQ(v-2\eta)
\label{funeqn}
\end{equation}
where
\begin{equation}
h(v)=\Theta(0)\Theta(v) H(v).
\end{equation}
and $\Theta(v)$ and $H(v)$ are the standard Jacobi theta functions.
It is important to recognize that the matrix $Q(v)$ in 
(\ref{funeqn}) is not unique. 
The original $Q$ matrix was discovered \cite{bax72} in 1972 under the
condition (\ref{root1}) and has been
further studied in detail in  \cite{Newdev}.
In 1973  Baxter constructed \cite{bax731} a second  $Q$-matrix which is
different from that of \cite{bax72} and is defined for generic values
of the crossing parameter
$\eta$. To distinguish between these two rather different
matrices we denote them  as  $Q_{72}$ and $Q_{73}$. 
Both matrices have the form
\begin{equation}
Q(v) = Q_R(v)Q_R^{-1}(v_0)
\label{QQinv}
\end{equation}
where $Q_{R,72}$ is only defined when (\ref{root1}) holds. 
Like the transfer-matrix $T(v)$ the matrix $Q_{R}(v)$ is 
the trace of a product of local matrices which are
in this case of size $L\times L$
\begin{equation}
[Q_R(v)]_{\alpha |\beta}={\rm Tr}S_R(\alpha_1, \beta_1)
S_R(\alpha_2, \beta_2)\cdots S_R(\alpha_N, \beta_N)
\label{TrSR}
\end{equation}
where $\alpha_j$ and $\beta_j=\pm 1$ and for the case $m_2=0$ in
(\ref{root1})
\begin{equation}
\begin{array}{lllll}
S_R(\alpha,\beta)(v)_{k,l} =& \delta_{k+1,l}& { u^{\alpha}(v+K-2k\eta)\tau_{-k,\beta}}&
                                    +\delta_{k,l+1}& { u^{\alpha}(v+K+2l\eta)\tau_{l,\beta}}+\\
                                    &\delta_{k,1}\delta_{l,1}&{ u^{\alpha}(v+K)\tau_{0,\beta}}&
                                    +\delta_{k,L}\delta_{l,L}&{ u^{\alpha}(v+K+2L\eta)\tau_{L,\beta}}\\
\end{array}
\label{SR}
\end{equation}
for $1<k\leq L$, $1<l\leq L$ and
where
\begin{equation}
u^{+}(v) = H(v) \hspace{0.5 in} u^{-}(v) = \Theta(v)
\label{ualpha}
\end{equation}

We have shown in \cite{Newdev} that $Q_{R,72}(v)$ has the 
(quasi)periodicity properties
\begin{equation}
Q_{R,72}(v+2K)= SQ_{R,72}(v)
\label{QRper}
\end{equation}
where 
\[
 S = \sigma_3 \otimes \sigma_3 \otimes \cdots \otimes \sigma_3
\]
and 
\begin{equation}
Q_{R,72}(v+2iK') = q^{-N}\exp(\frac{-i\pi Nv}{K})Q_{R,72}(v)
\label{QRqper}
\end{equation}
$Q_{72}$ can be formed in accordance with (\ref{QQinv}) only if  $Q_{R,72}^{-1}$ exists.\\
Whether $Q_{R,72}^{-1}$ exists depends on the size $N$ of the spin chain. If $N$ is so large 
that exact complete strings could exist then as we found 
in \cite{Newdev} 
\begin{quote}
\em \mbox{     }$Q_{R,72}^{-1}$ does NOT EXIST for real $\eta$ if
$m_1$ is even and $L$ is odd.
\end{quote}
\begin{quote}
\em
It exists only for odd $m_1$ and even or odd $L$.
\end{quote}
It follows that for odd $m_1$ and even or odd $L$ the matrix $Q_{72}$ exits.
It can be shown that it commutes with $T$. From (\ref{QRper}) and (\ref{QRqper}) 
it follows that its root structure differs from that of $Q_{73}$.
It is shown in \cite{Newdev} that
\begin{eqnarray}
Q_{72}(v)=\hat{{\cal K}}(q;v_k){\rm exp}(i(n_B-\nu)\pi v/2K)\prod_{j=1}^{n_B}
h(v-v^B_j)\nonumber\\
\times\prod_{j=1}^{n_L}H(v-iw_j)H(v-iw_j-2\eta)\cdots H(v-iw_j-2(L-1)\eta)
\label{form2}
\end{eqnarray}
\begin{equation}
2n_B+Ln_L=N.
\label{norootsa}
\end{equation}
$n_B$ is the number of Bethe roots $v_k$ and $n_L$ the number 
of exact $Q$-strings of length $L$.\\
{\it Note that $n_L$ is always even.}\\  
$ Q_{72}$ satisfies a functional relation conjectured in \cite{Newdev} and 
proven for $L=2$ in \cite{RIMS}:\footnote{It is 
not sufficient to prove this functional 
equation in the subspace of $R$-invariant states \cite{BazMang}.
It is important to show its validity in the degenerate subspaces of $T$.}
\\
For $ N$ even and either $ L$ even or $ L$ and $ m_1$ odd  
\begin{eqnarray} 
& & e^{ -N\pi i v/2K} Q_{72}(v-iK')\nonumber\\
&=&A\sum_{l=0}^{L-1} h^N( v-(2l+1)\eta)
\frac{Q_{72}(v)}{Q_{72}(v-2l\eta)Q_{72}(v-2(l+1)\eta)}
\label{con}
\end{eqnarray}
where $  A$ is a normalizing  constant matrix independent of $v$ 
that commutes with $Q_{72}$. \\
(\ref{con}) determines the complete set of zeros of 
each eigenvalue of $Q_{72}$. 
So it delivers more information than Bethe' equations 
(which are contained in (\ref{con})).
It determines the regular roots as well as the  roots 
appearing in exact strings:
If there exists an eigenvalue of $Q_{72}$ having $n_L$ 
exact strings in its set of zeros
with string centers 
\begin{equation} 
v_{c_{1}},\cdots ,v_{c_{n_{L}}}
\end{equation}
then there exist eigenvalues of $Q_{72}$ having the same 
regular roots and strings with centers
\begin{equation} 
v_{c_1}+\epsilon_1 i K',\cdots ,v_{c_{n_L}}+\epsilon_{n_{L}} i K'
\end{equation}
for all $ 2^{n_L}$ sets
\begin{equation} 
\epsilon_i = 0,1,~~~~i=1,\cdots, n_L
\end{equation}
It follows \cite{elliptic} : The eigenspace of a degenerate 
eigenvalue of $T$ has the dimension\\
 \[
 2^{n_L}~~~~~ {\rm or}~~~~~  2^{n_L-1}
\]
{\em
It is the purpose of the following sections to construct all 
eigenvectors of $T$ in these
degenerate subspaces.\\
}
We conclude this section with two remarks:\\
1.\\
Recently a $Q$-matrix has been found \cite{NewQ} for 
even $N$ which has the same properties as
$Q_{72}$ and which exists for $\eta=2m_1K/L$ where $Q_{72}$ does not exist.\\
2. \\
The matrix $Q_R(u)$ defined by (\ref{TrSR})-(\ref{ualpha})
which is the definition given literally in  \cite{bax72} 
only  satisfies the $TQ_R$ equation (C22) of \cite{bax72} 
when (\ref{root1}) holds with $m_2=0$. To extend the working of \cite{bax72}
to the case $m_2\neq 0$ the definitions of all theta functions must
be modified as done in \cite{bax731} for the related root of unity
condition (\ref{root2}). The required  modification is
\begin{eqnarray}
{\tilde \Theta}(v)=\Theta(v){\rm exp}[i\pi m_2(v-K)^2/(8KL\eta)]\nonumber\\ 
{\tilde H}(v)=H(v){\rm exp}[i\pi m_2(v-K)^2/(8KL\eta)]
\end{eqnarray}
With the replacements in (\ref{ualpha})  of $\Theta(v)\rightarrow
{\tilde \Theta}(v)$ and $H(v)\rightarrow {\tilde H}(v)$
the matrix $Q_{72}(v)$ of (\ref{QQinv}) will satisfy the $TQ$ 
equation (\ref{funeqn}) whenever it exists for 
all $\eta$ which satisfy (\ref{root1}).\\

\subsection{The string-free eigenstates of $T$.}
\noindent
Here we state the important result derived in \cite{TakFadd}:
If the crossing parameter is restricted to (\ref{root2}) with $m_2=0$
and $N$ is even there are  eigenstates of the transfer matrix $T$
given by
\begin{equation}   
\Psi_{m}
=\sum_{l=0}^{L-1}e^{2\pi iml/L}
B_{l+1,l-1}({ \lambda_1})\cdots B_{l+n,l-n}({ \lambda_n})
  \Omega_{N}^{l-n} 
\label{TFstate}
\end{equation} 
where $ \lambda_1, \cdots , \lambda_n$ are chosen to satisfy
\begin{equation}  
\frac{h^{N}( \lambda_j+\eta)}{h^{N}( \lambda_j-\eta)}=
e^{-4\pi im/L}\prod_{k=1,k\neq j}^{n}
\frac{h(\lambda_j-\lambda_k+2\eta)}{h(\lambda_j-\lambda_k-2\eta)}.
\label{Betheq}
\end{equation}
with $ N=2n+$integer$\times L$ and 
\begin{equation}  
\Omega_{N}^{l}=\omega_1^{l}\otimes \cdots \otimes \omega_N^{l}
\label{vac}
\end{equation}
\begin{equation}   
\omega_n^{l}=
\left(\begin{array}{c c}
 {H(s+2(n+l)\eta-\eta)}\\
  \Theta({s+2(n+l)\eta-\eta})\\
\end{array} \right)
\label{localvac}
\end{equation}
We call the set $ \lambda_1, \cdots , \lambda_n$ which solves the
Bethe-equations (\ref{Betheq}) regular Bethe-roots.
The $B_{k,l}$-operators are elements of the gauge transformed monodromy matrix
\[  
{\cal{T}}_{k,l} = M^{-1}_{k}(\lambda) {\cal T(\lambda)} M_{l}(\lambda) =
\left(\begin{array}{c c}
 A_{k,l} &  B_{k,l} \\
 C_{k,l} &  D_{k,l} \\
\end{array} \right).
\]
As mentioned above  the state vectors (\ref{TFstate}) 
give all singlet states
as well as a subset of each degenerate multiplet of states.

\subsection{Eigenstates of $T$ with strings.}
\noindent
The remaining much larger set of eigenvectors of $T$ has been obtained 
in \cite{elliptic}. These eigenvectors are given by
\noindent
\begin{equation}  
\Psi = \sum_{l=0}^{L-1}\omega^{l} B^{L,1}_{l}({ \lambda_{c_1}})...B^{L,1}_{l}({\lambda_{c_m}})\prod_{m=L_s+1}^{n}
B_{l+m,l-m}({ \lambda_m})  \Omega_{N}^{l-n}
\label{psi1}
\end{equation}
where $\omega = \exp(2\pi i ml/L)$ and the contribution of an exact $L$-string is
\begin{equation} 
B^{L,1}_{l}(\lambda_c)=\sum_{j=1}^{L_s}B_{l+1,l-1}(\lambda_1)\cdots
\left(\frac{\partial B_{l+j,l-j}}{\partial \eta}(\lambda_{j})
-{\hat{Z}_j}\frac{\partial B_{l+j,l-j}}{ \partial \lambda}(\lambda_{j})\right)
\cdots B_{l+L_s,l-L_s}(\lambda_{L_s}) 
\label{stringop}
\end{equation}
and where $\lambda_{c_{i}}$ are the centers of strings
\begin{equation}
  \lambda_k = \lambda_c-2(k-1)\eta \hspace{0.6 in} k=1,\cdots,L_s
\label{string}
\end{equation}
We shall refer to (\ref{stringop}) as B-string operator which 
creates the B-string (\ref{string}).
The string centers are free parameters. The string length $L_s$ is for odd $L$
\[
L_s = L
\] 
and for even $L$
\[
L_s = L/2
\]
The key of our method is the function $\hat{Z}_j$ in (\ref{stringop}). Its definition is
\begin{equation}
\hat{Z}_{j}(\lambda_c)=\hat{Z}_1(\lambda_c-(j-1)2\eta)
\end{equation}
\begin{equation}
\hat{Z}_1=-2\frac{\sum_{k=0}^{L_s-1}k\frac{\omega^{-2(k+1)}\rho_{k+1}}{P_kP_{k+1}}}
{\sum_{k=0}^{L_s-1}\frac{\omega^{-2(k+1)}\rho_{k+1}}{P_kP_{k+1}}}
\label{X1}
\end{equation}
\begin{equation}
\rho_k=h^{N}(\lambda_c-(2k-1)\eta)
\label{rho}
\end{equation}
and 
\begin{equation}
P_k = \prod_{m=L_s+1}^{n}h(\lambda_c-\lambda_{m}-2k\eta).
\label{P}
\end{equation}
$ \omega=e^{2\pi i m/L}$ is a Lth root of unity.
The constants $\lambda_m$ in (\ref{P}) are the regular Bethe-roots of the state
under consideration. We see that each string operator depends on its string center
$\lambda_c$ , on $s,t$ and on all regular roots defining the eigenvalue of $T$.\\
We briefly describe how the result (\ref{psi1})-(\ref{P}) has been derived.
The attempt to generate the missing eigenstates by adding a complete exact string 
\begin{equation}
 \lambda_k
= \lambda_c-2(k-1)\eta,~~~k=1,\cdots L_{s} 
\label{r}
\end{equation}
to the set of Bethe roots fails because
\begin{equation}
 B_{l+1,l-1}(\lambda_1)\cdots B_{l+L_s,l-L_s}(\lambda_{L_s}) = 0
\label{BL}
\end{equation}
\noindent
To circumvent this problem we insert instead of (\ref{r}) the expression
{ 
\begin{equation}
 \lambda_k
= \lambda_c-2(k-1)\eta-\hat{Z}_k(\lambda_c)\epsilon,~~~k=1,\cdots L_{s} 
\label{Zr}
\end{equation}}
and perform the limit $\epsilon \rightarrow 0$. The $O(\epsilon )$ term 
gives the desired result.The function  $\hat{Z}_k(\lambda_c)$ is an essential
ingredient of our method.\\
We follow the same path as Takhtadzhan and Faddeev \cite{TakFadd}:
Let $ \Psi$ be the candidate for an eigenstate of $T$ and
\begin{equation}
 T\Psi  = \underbrace{{t(\lambda)} \Psi}_{\rm wanted~~term}
+ ~~ \underbrace{\rm additional~~terms}_{\rm unwanted}
\end{equation}
In the work of \cite{TakFadd}
the unwanted terms can be removed by adjusting the set of free parameters 
$\lambda_1, \cdots ,\lambda_n$ 
to satisfy '' Bethe's equations''.
In our case a first set of unwanted terms is removed 
by again invoking Bethe's equations
to determine $\lambda_1 ,\cdots \lambda_n$.\\
The remaining second set of unwanted terms is removed 
by appropriately choosing 
the functions $\hat{Z}_k(\lambda_c)$ which leads to 
\begin{eqnarray}
&& \omega(\hat{Z}_{k+1}-\hat{Z}_{k}-2)\rho_{k-1}P_{k} = 
 \omega^{-1}(\hat{Z}_{k}-\hat{Z}_{k-1}-2)\rho_{k}P_{k-2}
\label{XX}
\end{eqnarray}
where $\rho$ and $P_k$ and are defined in (\ref{rho}) and (\ref{P}). 
The solution of (\ref{XX}) is
(\ref{X1}).\\
The total number of operators $B$ building an eigenvector is restricted by
\begin{equation}
2(n_B+n_s) + rL = N
\end{equation}
where $n_B$ is the number of regular roots and $n_s$ is the number of roots
belonging to $B$-strings. It is important to note that the 
integer $r$ may be positive
and negative. It follows that there is no restriction on the number of 
$B$-string-operators in a state vector. To understand the role of B-strings we note
that the analytical expression for eigenstates of the eight-vertex model without strings
given by equ. (\ref{TFstate}) 
depends on two free parameters $s,t$. For degenerate eigenstates which form a 
space of dimension $d$ a subspace of dimension $d_0 < d$
can be constructed by the variation of $s$ and $t$ without applying B-string operators
(\ref{stringop}) . Detailed numerical studies have revealed that 
the variation of $s,t$ will give only for very small d (e.g. $d$=2)
the full degenerate eigenspace. In all other cases one has to add B-string
operators with the additional freedom to choose the string center 
to generate the full subspace. After this is achieved by adding a certain number of 
B-strings (the exact number depends on the system size $N$ and the value of $\eta$) 
the addition of more and more B-strings will only map this subspace into itself.
In particular adding B-strings to a singlet state with
$n_B=N/2$ Bethe-roots does not destroy this state but reproduces it.\\
It is obvious that the B-string operators (\ref{stringop}) play the role of
symmetry operators in the eigenspaces of the transfer matrix. As they generate
for each eigenvalue the full degenerate subspace they must contain all information about
the hidden symmetry algebra. The properties described above raise the possibility
that the B-string operators (\ref{stringop}) are related to elliptic current operators
in cyclic representations. We expect that the $q \rightarrow 0$ limit of this algebra is
the $sl_2$-loop algebra which has been found in \cite{dfm} and \cite{Odyssey} to 
describe the symmetry of the six-vertex model. This suggests that the symmetry
is related to $U_q(\widehat{sl_2})$. To our knowledge the theory of cyclic 
representations of $U_q(\widehat{sl_2})$ is still undeveloped.
\footnote{ We thank V. Chari, A. Kirillov and V. Kac for this information.}

\centerline{\bf \large Acknowledgements}
\noindent
We wish to thank the organizers of this Solvay conference for the
opportunity to participate in the celebration of the 75th anniversary
of Bethe's ansatz.

\end{document}